\documentclass[prd,twocolumn,floats,floatfix,nofootinbib,prd,tightenlines,superscriptaddress]{revtex4}
\usepackage{axodraw}
\usepackage{graphicx}
\usepackage{multirow}
\usepackage{epstopdf}
\usepackage{amsmath}    % need for subequations
\usepackage{color}
\def\be{\begin{equation}}
\def\ee{\end{equation}}
\def\bea{\begin{eqnarray}}
\def\eea{\end{eqnarray}}

\newcommand{\gsim}{ \mathop{}_{\textstyle \sim}^{\textstyle >} }
\newcommand{\lsim}{ \mathop{}_{\textstyle \sim}^{\textstyle <} }

\newcommand{\missp}{{p}_T \!\!\!\!\!\! /\;}
\newcommand{\misse}{{E}_T \!\!\!\!\!\!\! /\;\;}
\newcommand{\missh}{{H}_T \!\!\!\!\!\!\!\! /\;\;}
\newcommand{\miss}[1]{{#1} \!\!\! /}

\begin{document}

\title{Dijet Searches for Supersymmetry at the LHC}

\author{Lisa Randall}
\affiliation{Jefferson Physical Laboratory \\Harvard University, Cambridge, MA 02138}
\author{David Tucker-Smith}
\affiliation{Department of Physics,Williams College,\\ Williamstown, MA 01267}

\begin{abstract}
We present several strategies for searching for supersymmetry in dijet channels that do not explicitly invoke missing energy. Preliminary investigations suggest that signal-to-background ratios of at least 4--5 should be achievable at the LHC, with discovery possible  for squarks as heavy as $\sim$ 1.7 TeV.

\end{abstract}

\maketitle

\section{Introduction}
The LHC is set to explore the physics of the weak scale, whatever it
should turn out to be. Supersymmetry is one of the leading candidates and enormous effort has been dedicated to studying missing energy signals that characterize almost any weak-scale
supersymmetric model.
 However, supersymmetry searches will be challenging and disentangling the
supersymmetry parameters will be  more difficult still.

In light of the above, it is imperative to study every possible channel in
order to optimize our chances of discovering {new physics} and understanding the
underlying theory.  In this regard, events with the lowest multiplicity {may} be the simplest {ones} with which to make headway on the inverse problem.

Although two-jet events with missing energy have been studied at the
Tevatron  \cite{:2007ww}, they have been less prominent in LHC studies.
 ATLAS has shown that two jet events can be useful for certain SUSY models,
 both for discovery and for constraining superpartner masses \cite{atlas},
 but recent ATLAS and CMS studies have focused more heavily on the more
challenging cascade decays.  In this paper we study one novel and two existing kinematic variables that can be used to capture dijet missing-energy events without explicit reference to missing transverse energy.  We find that pairs of these variables can be used to give signal-to-background of at least 4--5, indicating that these variables are worth exploring with a full detector simulation\footnote{Such a study has been started by CMS \cite{cms}.  In addition,  ATLAS is currently engaged in an updated dijet study \cite{ian}.}.

Dijet events are worthy of attention as a potentially clear window into parameter space. They
are not results of complicated cascade decays but arise simply from two squarks
decaying to two quarks and two neutralinos. Because we know the identity
of the particles involved and because there are so few, the signal is relatively straightforward
to interpret. For example, with sufficient integrated luminosity, these events alone can be used to constrain the squark and neutralino masses. Dijet studies along the lines explored here may usefully supplement recent analyses dedicated to distinguishing SUSY from other models using events with at least three jets \cite{Hubisz:2008gg}.

The kinematic variables we consider are constructed from the two jets' momenta alone.
These variables should  have different systematic uncertainties than missing transverse energy  since they pick out  slightly different events and are based on different measurements.  At the very least, then, the searches we suggest should be worthwhile as cross-checks of standard searches.  The variables we use may also be useful for optimization when signal-to-background is relatively low.

The searches we describe will be most effective  when squarks are pair-produced in abundance and have large branching ratios to decay directly to the lightest neutralino, which requires that squarks are lighter than the gluino so that cascade decays through gluinos  are absent.  Because $t$-channel gluino exchange is an important source of squark pair production, the lighter the gluino the more prominent the signal.  For the parameter points considered below, we find the signal is cut by a factor of $\sim 6-7$ when the gluino decouples. Fortunately, comparable gluino and squark masses are  a feature of 
  a large class of models -- most notably
high-scale models where the heavier gluino mass feeds into the squark
mass.  We focus on such models in this study.

\section{Analysis Details}

Before getting to the dijet properties that will be the focus of our study, we consider the effectiveness of $\misse\!\!$ and  $\missh$, the missing transverse energy obtained from the dijet system alone. After requiring the sum of the two jets' $p_T$'s to be greater than 500 GeV,  event rates and signal-to-background ratios for one particular SUSY point  are presented in Table \ref{table:met}  (details regarding event generation and cuts  are given below).   Neither variable suffices for a clean search, but we observe that the $S/B$  values obtained using  $\missh$ are essentially identical  to those obtained using $\misse$.  This analysis suggests that, in the two-jet channel at high $p_T$, nothing is to be gained by using full $\misse$ rather than kinematic variables associated with the two jets alone.
 %%%%%%%%%%%%%%%% table with missing energy and mht%%%%%%%%%%%%%%%%%%% 
 %%%%%%%%%%%%%%%%%%%%%%%%%%%%%%%%%%%%%%%%%%%%%%%%%%% 
 \begingroup
\squeezetable
\begin{table}[t]
\begin{center}
\begin{tabular}{||c||c|c|c|c|c|c|c|c|c|c||}
\hline\hline
& $\misse/\missh $ cut  & 300 & 350 & 400 & 450 & 500 & 550 & 600 &  650 & 700  \\
\hline\hline
    \multirow{2}{*}{$\misse$} & $\sigma_{susy}$(fb) &   864. & 759. & 645. & 526. & 397. &
   257. & 143. & 81.9 & 51.1 \\
   \cline{2-11}
 & $S/B$  &  0.7 & 1.0 & 1.3 & 1.7 & 1.8 & 2.0 & 1.8 & 1.5 & 1.4\\
    \hline \hline
    
    \multirow{2}{*}{$\missh$} & $\sigma_{susy}$(fb)  & 862. & 757. & 639. & 521. & 379. &
   229. & 128. & 74.5 & 47.4 \\
   \cline{2-11}
 &  $S/B$ & 0.7 & 1.0 & 1.3 & 1.7 & 1.9 & 1.8 & 1.7 & 1.5 & 1.3 \\
\hline\hline

\end{tabular}
\end{center}
\label{default}
\caption{For dijet events passing the cuts described in the text, the dependence of the signal cross section and  signal-to-background ($S/B$) on
a variable $\misse$ cut (top), and on a variable $\missh$ cut (bottom).  All energies are in GeV.}
\label{table:met}
\end{table}%
 \endgroup
 %%%%%%%%%%%%%%%%%%%%%%%%%%%%%%%%%%%%%%%%%%%%%%%%%%%

We now present three dijet variables   that can be used to separate signal and bacground, with $\sim 1$\% of signal events passing all cuts.

\begin{itemize}
{\item $\alpha$:} which we define as the ratio of the  $p_T$ of the second hardest jet and the invariant mass formed from the two hardest jets,
\begin{equation}
\alpha \equiv \frac{{p_T}_2}{m_{jj}}.
\end{equation}
As far as we know, this variable has not been considered  previously.  
Background events generally trail off at $\alpha=0.5$, whereas supersymmetry events with invisible decay products can easily have larger $\alpha$.
Large $\alpha$ tends to arise in  events in which the jets are not back-to-back.  As one extreme example, if the two jets are nearly aligned, their invariant mass can be quite small, leading to very large $\alpha$.

Because of the  background's sharp drop-off  around $\alpha=0.5$, this variable is potentially useful as a  diagnostic tool  for analyzing two jet events and cleanly separating signal events from QCD.

{\item $\Delta \phi $:} the azimuthal angle between the two hardest jets. Azimuthal angle is often used in conjunction with missing transverse energy, and  $\Delta \phi $ was among the variables used in the dijet SUSY search at D0 \cite{:2007ww}.

{\item $M_{T2}$ \cite{Lester:1999tx}:}  which is defined  for events in which two particles of the same mass undergo identical semi-invisible decays, as
 \begin{equation}
 M_{T2}(\chi) = \min_{\miss{q}_1+\miss{q}_2=\miss{p}_T} \{ \max[m_T(p_1, q_1\!\!\!\!\! / \;,\chi), m_T(p_2,q_2\!\!\!\!\! / \;,\chi)] \},
 \end{equation}
 where $p_1$ and $p_2$ are the momenta of the visible particles, $\missp$ is the missing transverse momentum of  the event, and $m_T$ is the transverse mass function, which depends on an assumed value $\chi$ of the invisible particle's mass. In calculating  $M_{T2}(\chi)$ we use the missing transverse momentum as determined by the dijet system alone.

 If $\chi$ is taken to be equal to the mass of the invisible particle, the $M_{T2}$ distribution will have an endpoint at the mass of the decaying particle.  
 Not knowing this mass, $M_{T2}$ endpoints still constrain the masses of the decaying and invisible particles, as emphasized in \cite{Lester:1999tx} and used below.

\end{itemize}
%%%%%%%%%%%%%%%%%%%%%%%%%%%%%%%%%%%%%%%%%%%%%%%
\begin{figure*}[t] %  figure placement: here, top, bottom, or page
   \centering
   \includegraphics[width=2in]{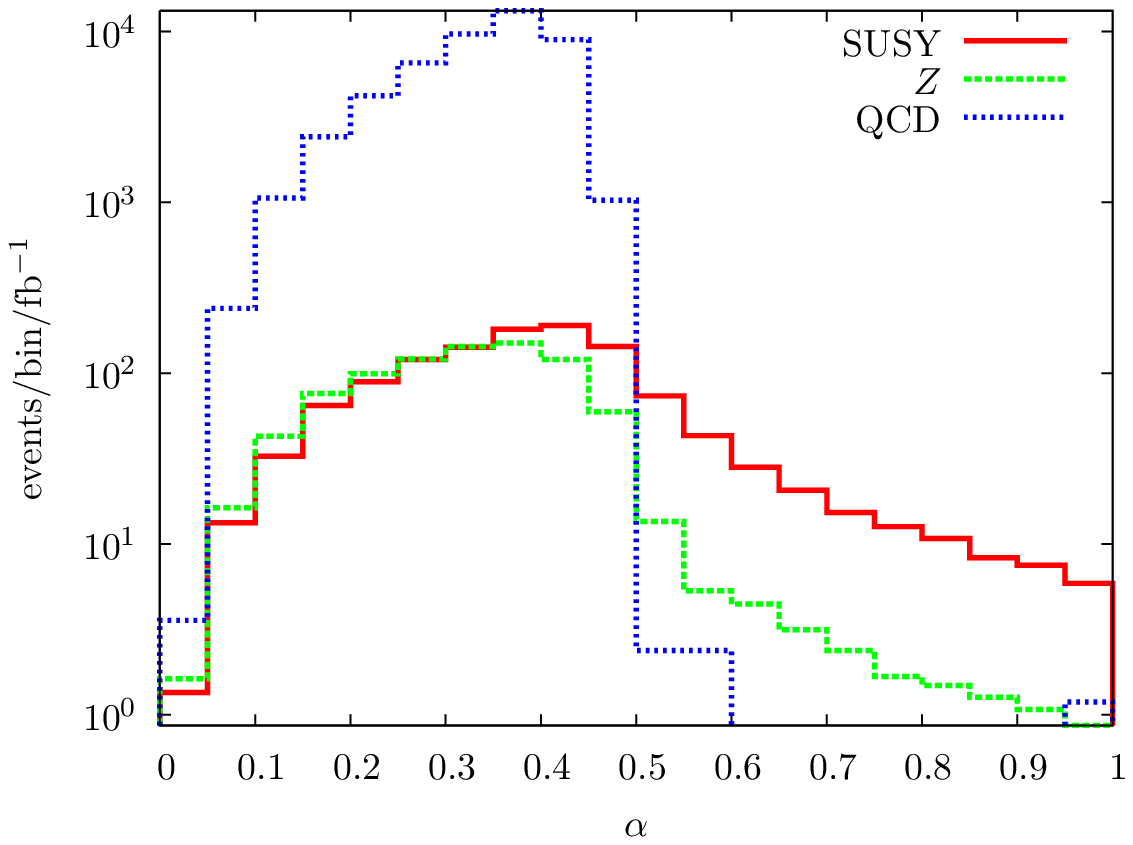} \quad\quad
      \includegraphics[width=2in]{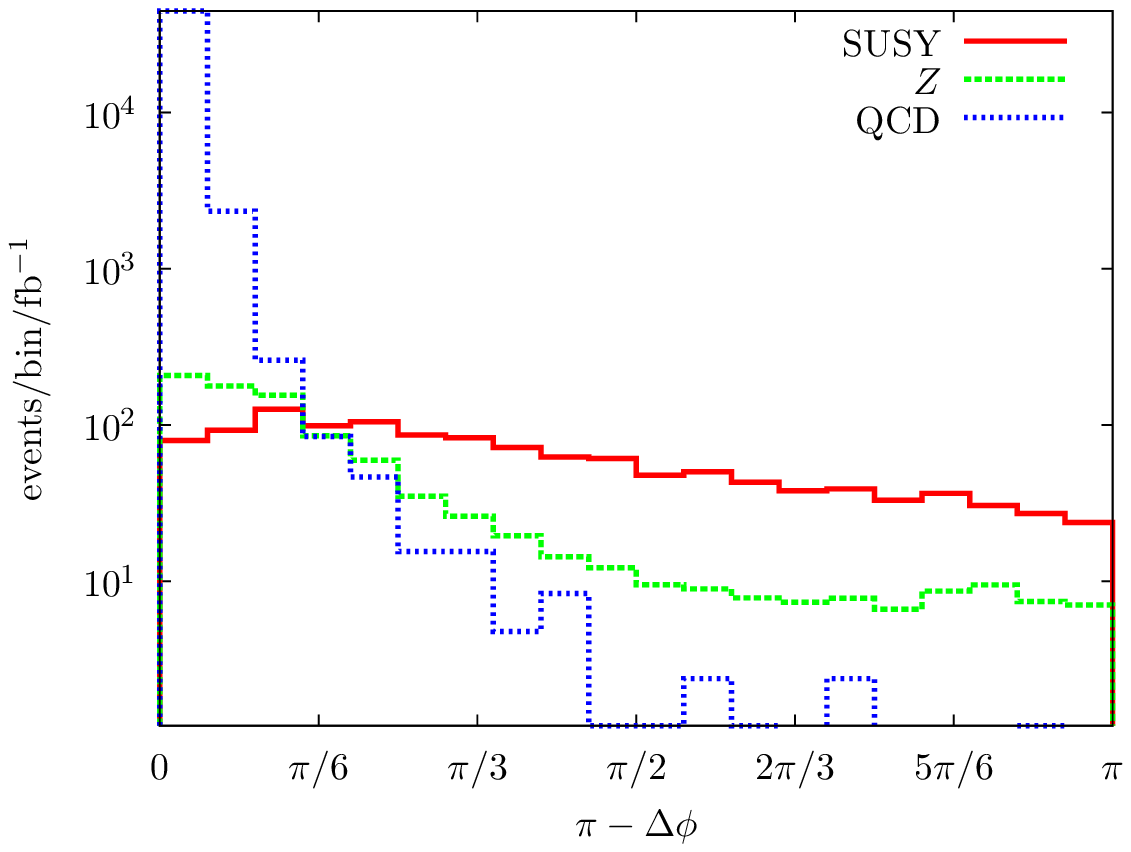}  \quad\quad
         \includegraphics[width=2in]{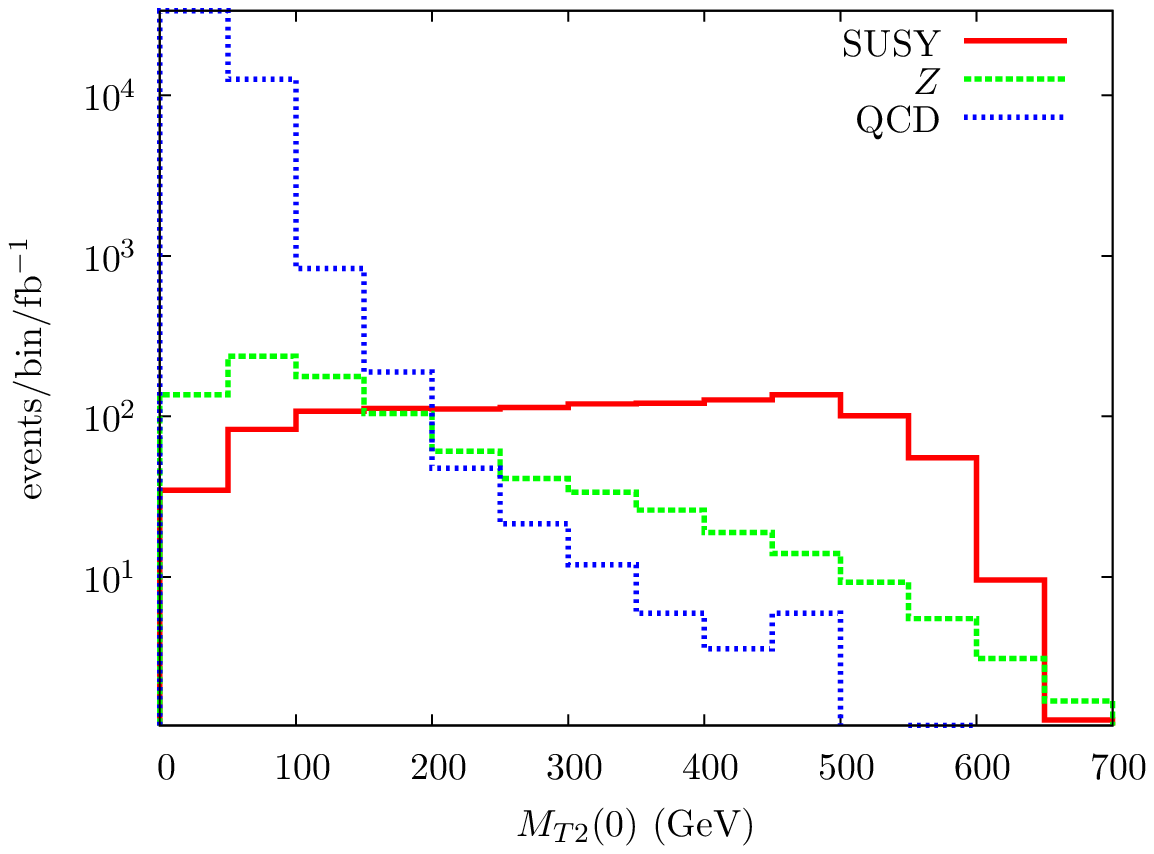}
   \caption{SUSY, $Z \rightarrow \nu {\overline \nu}$+jets and QCD rates for passing the cuts described in the text, as functions of  $\alpha$ (left), $\Delta \phi$ (middle), and  $M_{T2}(0)$ (right). }
   \label{fig:phialphascans}
\end{figure*}
%%%%%%%%%%%%%%%%%%%%%%%%%%%%%%%%%%%%%%%%%%%%%%%

We consider these variables singly and in tandem. We find the first two variables are useful in that one can choose parameter-independent cuts that give sizable $S/B$, whereas the last variable, though more parameter-dependent in its optimization, might ultimately maximize $S/B$. Since the advantage is not overwhelming, we  expect all the variables could prove  useful, either at the trigger or analysis level.  Because they are dimensionless, the first two variables might have the further advantage of being less sensitive to absolute energy scale, and might therefore have lower systematic errors.

For all  our analyses, we select events in which exactly two jets have $p_T>50$~GeV, with no isolated leptons, photons, or $\tau$ jets.  One could attempt to achieve better background rejection by an additional veto on extra jets with lower $p_T$.  In general, we have chosen felicitous cuts but have not pursued a careful optimization, which will be more appropriate at the full-detector-simulation level.

A  gluino that  is only slightly  heavier than the squarks arises  naturally in models with supersymmetry broken at a high scale, as renomalization-group effects prevent the squarks from being hierarchically lighter than the gluino.  For our analyses we specify parameters at the high scale and use the SUSY-HIT package \cite{Djouadi:2006bz} to calculate superpartner masses and decay branching ratios. In the relevant parameter regions, the signal depends strongly on $M_{1/2}$, the unified gaugino mass at the high scale, and  is less sensitive to $M_0$, the unified scalar mass, because the squark mass is dominated by gauge-loop contributions.   We set the other SUSY parameters to be $\tan \beta = 10$, $A_0=0$, and $\mu>0$.

The backgrounds included in our analyses are QCD, $(W\rightarrow l {\overline \nu})/(Z\rightarrow \nu {\overline \nu})$+jets, and ${t \overline t}$.   We have checked that diboson+jets production does not significantly modify our results.  The  QCD and $t {\overline t}$ samples were generated with Pythia 6.4 \cite{Sjostrand:2006za}, and $Z/W$+jets with Alpgen 2.12 \cite{Mangano:2002ea}. Fully showered and hadronized events were then passed to the PGS 4.0 detector simulator \cite{PGS}, with the energy smearing in the hadronic calorimeter given by  $\Delta E/E = 0.8/\sqrt{E/{\rm GeV}}$ and the calorimeter granularity set to $(\Delta \phi\times \Delta \eta)= (0.1\times 0.1)$. Jets  were defined using a cone algorithm with $\Delta R=0.4$.    

A $K$-factor of 2 is applied to the QCD sample, but no $K$-factor is used for  $W/Z$ production, because the most important contributions  come from  $W/Z$+2 jets, which  are not enhanced at NLO \cite{Campbell:2003hd}.  (After cuts, $W/Z$ production ends up being  the dominant background to SUSY dijet events, so to include a $K$-factor one can simply divide our signal-to-background ratios by $K$.)  For $t{\overline t}$ we use $\sigma=830$ pb as the NLO  production cross section \cite{Bonciani:1998vc}.  Including  the $K$ factors our samples sizes are  $ \sim 0.8$ fb$^{-1}$ for QCD, $\sim$ 20 fb$^{-1}$ for $t{\overline t}$, and $\sim 100$ fb$^{-1}$  for $W/Z$.  Appropriate generator-level kinematic cuts were imposed to obtain the QCD and $W/Z$ samples.  

SUSY samples were also generated with Pythia.  For each parameter point  we use Prospino 2.0 \cite{prospino} to calculate an appropriate $K$-factor from the NLO cross section for squark pair production \cite{Beenakker:1996ch}.

\section{Results}

\begin{figure}[t] %  figure placement: here, top, bottom, or page
   \centering
   \includegraphics[width=3in]{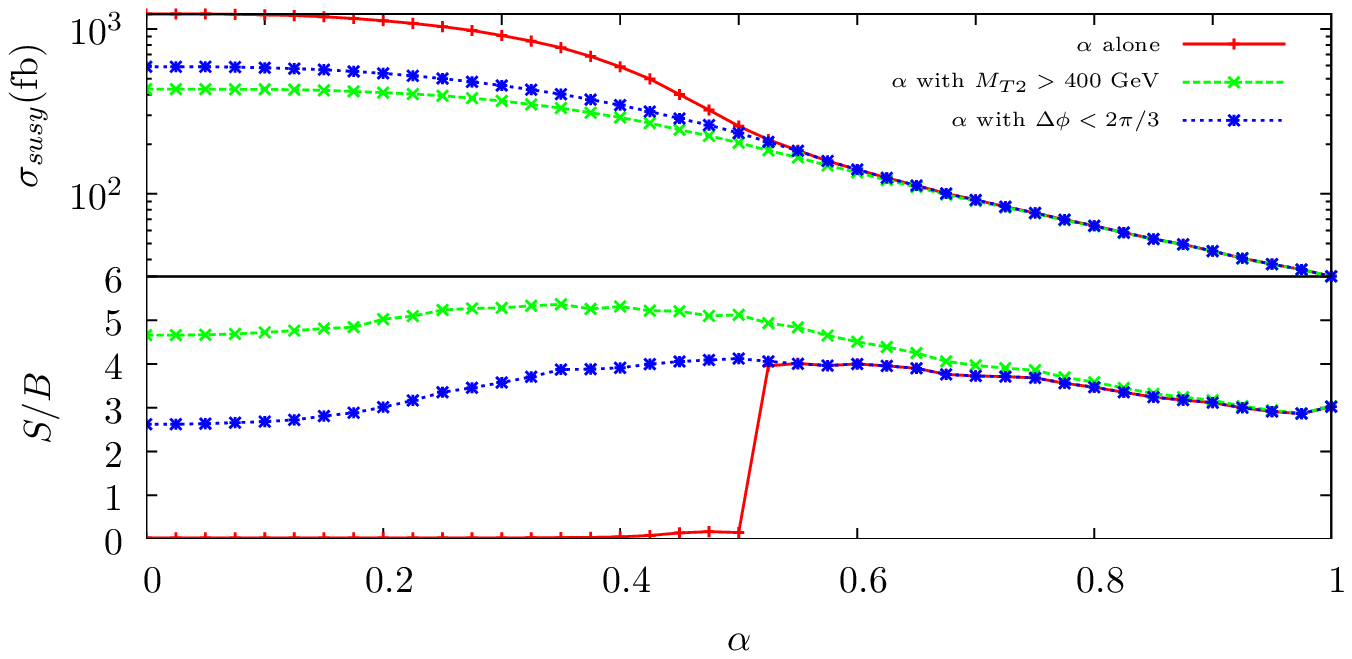} \\
   \vspace{0.2cm}
   
      \includegraphics[width=3in]{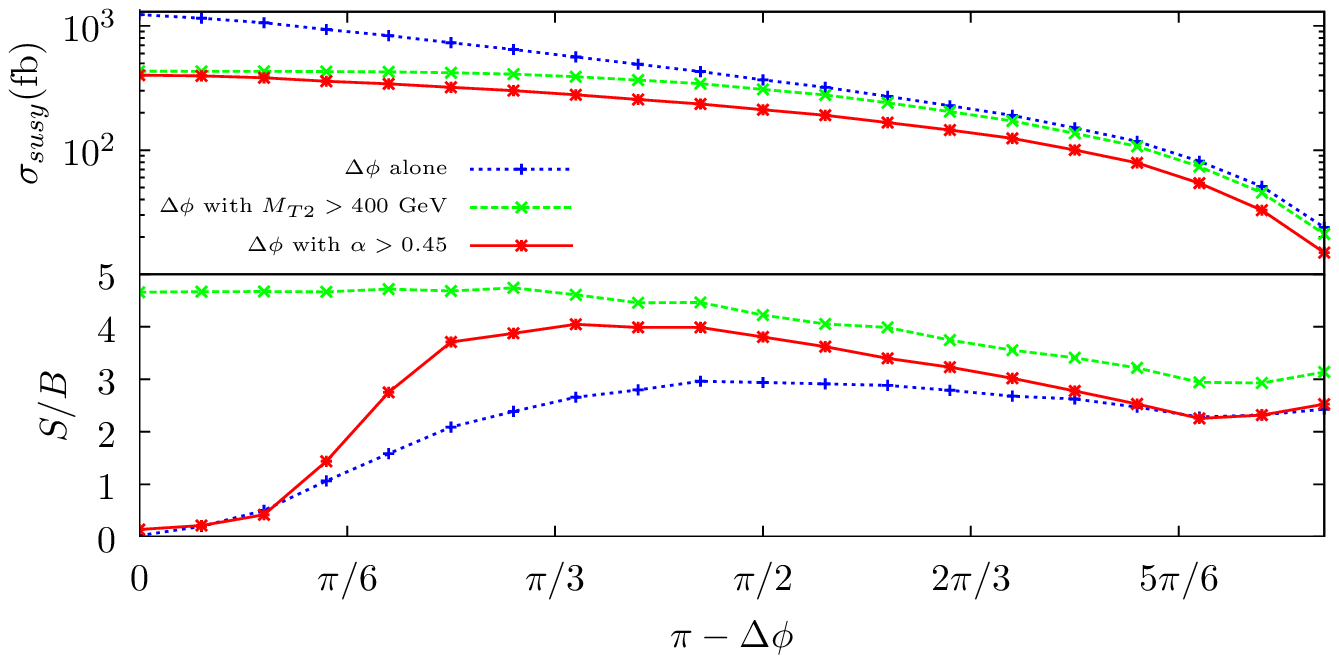}  \\
     \vspace{0.2cm}

         \includegraphics[width=3in]{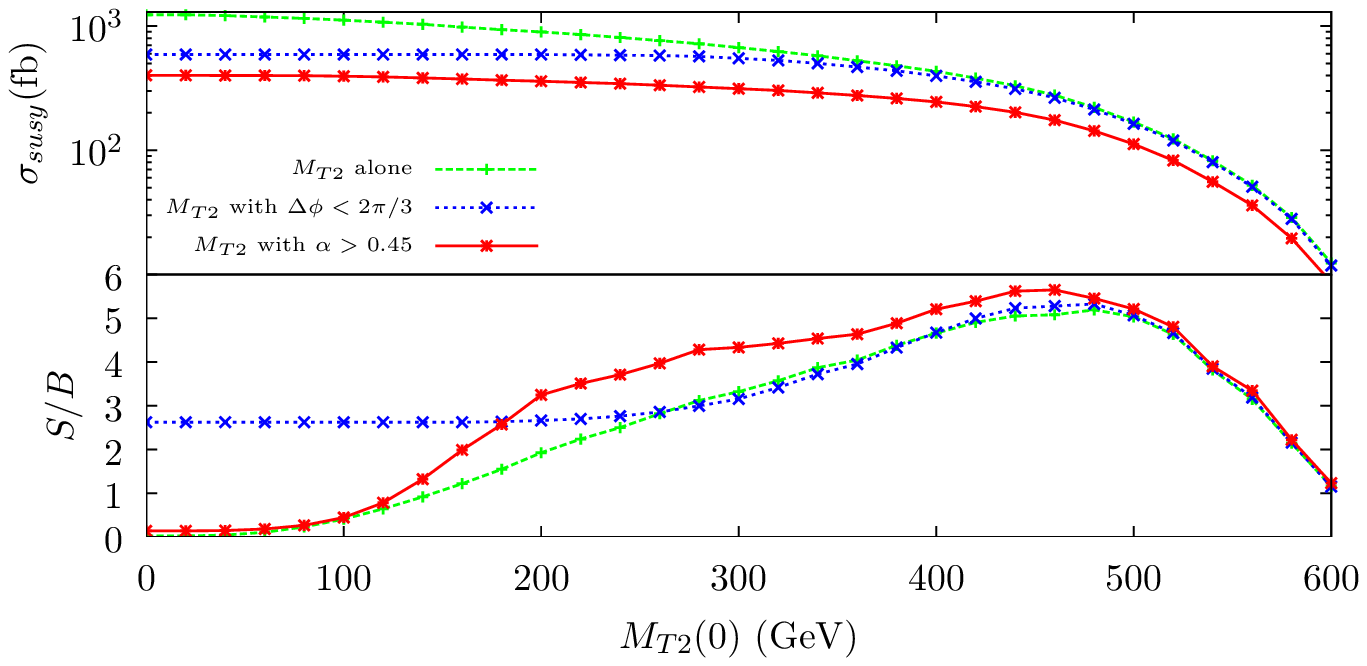}
   \caption{For events preselected as described in the text, the dependence of the signal cross section and  $S/B$ on a variable  $\alpha$ cut (top),  a variable  $\Delta \phi$ cut (middle), and a variable $M_{T2}$ cut (bottom). 
   }
   \label{fig:nomet}
\end{figure}

The plots in Figure \ref{fig:phialphascans} suggest that appropriate cuts on   $\alpha$,  $\Delta \phi$, and/or $M_{T2}$ can suppress both the QCD background and the dominant background after cuts,  $(Z \rightarrow \nu {\overline \nu})$+jets. The SUSY parameter point used here is $(M_{1/2},M_0) = (300,100)$~GeV, and we impose a  hard cut on the sum of the two hard jets' transverse momenta,
\begin{equation}
{p_T}_1+{p_T}_2 > 500 \;{\rm GeV}.
\end{equation}
To streamline the analysis, events were required to have $\misse> 100$ GeV for Figure, \ref{fig:phialphascans} and at least one of
 $\alpha > 0.5$, $\Delta \phi <2\pi/3$, and $\misse>100$ GeV for Figure  \ref{fig:nomet}.  Removing these requirements does not affect the results once optimal cuts on  $\alpha$,  $\Delta \phi$, and/or $M_{T2}$ are made. 

Evidently signal dominates over background for $\alpha\gsim 0.5$, $\Delta \phi \lsim 2\pi/3$, and $M_{T2} \gsim 300$ GeV.
We will soon see that $\alpha $, $\Delta \phi$, and $M_{T2}$  can be used to discriminate signal from 
background by themselves, but first we point out that 
 cuts on these variables can improve an analysis based on $\misse\!$ or $\missh\!$.  For example, the combination $(\alpha>0.45,\; \missh\!>300 \:{\rm GeV})$ selects 315 signal events per fb$^{-1}$, with $S/B=4.3$.   The combination $(\Delta \phi < 2\pi/3,\; \missh\!>450 \:{\rm GeV})$ gives a somewhat lower $S/B$ (3.1), but with more events (429).    An $M_{T2}$ cut of 450 GeV gives the largest $S/B$ of all (5.0, with 304 events), and in fact there appears to be no benefit in supplementing the $M_{T2}$ cut with the $\missh$ cut.  

Figure \ref{fig:nomet} suggests 
%%%%%%%%%%%%%%%%%%%%%%%%%%%%%%%%%%%%%%%%%%%%%%%%%%%
%%%%%%%%%%%%%%%%%%%%%%%%%%%%%%%%%%%%%%%%%%%%%%%%%%%
that each of  $\alpha$, $\Delta \phi$, and $M_{T2}$ can be used  independently  to observe a clear signal, without employing $\missh$ at all. Well-chosen cuts give $\sim {\rm  a \;few} \times 10^2$  signal events after 1 fb$^{-1}$, with $S/B\sim 3-5$.  

Figure \ref{fig:nomet}  also shows how  the three variables can be used in pairs to improve $S/B$ in conjunction with the signal event-rate.  We again find that  $M_{T2}$ seems to dominate a little, but since we do not know if this is the cleanest variable to use in practice, which  can be determined nly after a full detector simulation, we present all combinations. Any two on their own can potentially give a robust signal.

As an  exanple, we consider the combination $\Delta \phi < 2\pi/3$ and  $\alpha< 0.45$, which gives a good $S/B$ and a decent event rate. As stated earlier, we do not optimize cuts, but we use this combination that works rather well.

\begin{figure}[h] %  figure placement: here, top, bottom, or page
   \centering
   \includegraphics[width=3in]{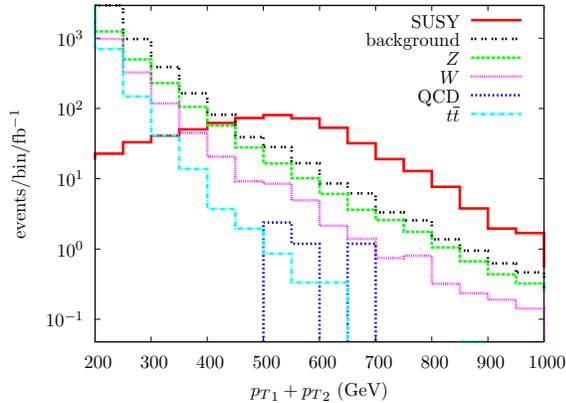} 
   \caption{Signal and background rates after the cuts $\Delta \phi < 2\pi/3$ and $\alpha >0.45$.  The QCD background is not included for ${p_1}_T +{p_2}_T<500$ GeV.   We take $(M_{1/2}, M_0)=(300,100)$ GeV. }
   \label{fig:300ptbins}
\end{figure}

With those cuts in place,  Figure \ref{fig:300ptbins} shows signal and background events binned in the sum of the two hardest jets' transverse momenta.
%%%%%%%%%%%%%%%%%%%%%%%%%%%%%%%%%%%%%%%%%%%%%%%%%%%%%%%%%%%%%%%%%%%%%%%%%%%%%%%%%%%%%
%%%%%%%%%%%%%%%%%%%%%%%%%%%%%%%%%%%%%%%%%%%%%%%%%%%%%%%%%%%%%%%%%%%%%%%%%%%%%%%%%%%%%
We see that $Z$+jets is the dominant background, followed by $W$+jets.  A total of four QCD events with ${p_1}_T +{p_2}_T<500$ GeV passed the cuts, out of a sample corresponding to over 1.5 fb$^{-1}$ of integrated luminosity, divided by the $K$ factor.  A higher luminosity sample would be needed to get a better estimate of the QCD background, but it seems safe to say that the $W$ and $Z$ backgrounds are more important.

In Figure \ref{fig:300ptbins}  we see that  $S/B$ is cleanest at high $p_T$. Of course the optimal $p_T$  cut depends on  underlying parameters that are not known {\it a priori}, but a scan at high $p_T$ should help maximize $S/B$. For the chosen parameter point, cutting above ${p_1}_T +{p_2}_T=550$ GeV gives  $S/B = 4.9$, with an average of 205  signal events after 1 fb$^{-1}$. Table \ref{table:efficiencies} shows the efficiencies with which the SUSY events pass the successive jet multiplicity, ${p_T}_1+{p_T}_2$, $\Delta \phi$, and $\alpha$ cuts.
%%%%%%%%%%%%%%%%%%%%%%%%%%%%%%%%%%%%%%%%%%%%%%%%%%%%%%%%%%%%%%%%%%%%%%%%%%%%%%%%%%%%%
\begingroup
\squeezetable
\begin{table}[t]
\begin{center}
\begin{tabular}{||c||c|c|c|c||}
\hline \hline
& $N_{jets}=2$ & ${p_T}_1+{p_T}_2 > 550$ & $\Delta \phi < 2\pi/3$ & $\alpha>0.45$  \\ \hline \hline
$\epsilon$&$1.08 \times 10^{-1}$&$5.04 \times 10^{-2}$ & $2.05 \times 10^{-2}$& $9.48 \times 10^{-3}$  \\
\hline
$\sigma_{susy}$ (fb) & $2.33 \times 10^{3}$&  $1.09 \times 10^{3}$ & 443. & 205. \\\hline \hline
\end{tabular}
\end{center}
\label{default}
\caption{The efficiencies $\epsilon$ for signal events to pass the successive cuts, taking $(M_{1/2}, M_0)=(300,100)$ GeV. }
\label{table:efficiencies}
\end{table}%
\endgroup
%%%%%%%%%%%%%%%%%%%%%%%%%%%%%%%%%%%%%%%%%%%%%%%%%%%%%%%%%%%%%%%%%%%%%%%%%%%%%%%%%%%%%

 The final efficiency is lower than that for SUSY searches with  additional jets, and so  despite the different systematics SUSY might well be discovered in other channels first.  Moreover, the dijet channel is relevant only for certain models.  On the other hand, this analysis picks out  particularly simple events-- two squarks decay to produce two jets and two neutralinos.  If these events do occur it would certainly be worthwhile to study them in isolation.  
 
 For example, with enough luminosity these events alone can be used to obtain a simple constraint on the squark and neutralino masses, using the $M_{T2}$ event function \cite{Lester:1999tx} introduced above. If one can ignore all visible particles in the event except those in the two jets, one expects %%%%%%%%%%%%%%%%%%%%%%%%%%%%%%%%%%%%%%%%%%%%%%%%%%%%%%%%%%%%%%%%%%%%%%%%%%%%%%%%%%%%%
  \begin{figure}[h] %  figure placement: here, top, bottom, or page
   \centering
   \includegraphics[width=3in]{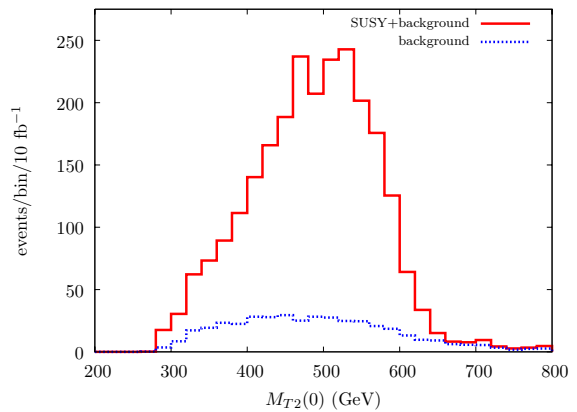} 
   \caption{The $m_{T2}$ distribution for signal and background, after the cuts described in the text.  We take $(M_{1/2}, M_0)=(300,100)$ GeV. }
   \label{fig:mt2}
\end{figure}
%%%%%%%%%%%%%%%%%%%%%%%%%%%%%%%%%%%%%%%%%%%%%%%%%%%%%%%%%%%%%%%%%%%%%%%%%%%%%%%%%%%%%
 the endpoint
 \begin{equation}
 M_{T2}(0)_{max}=\frac{m_{\tilde q}^2-m_{\tilde \chi_1^0}^2}{m_{\tilde q}}.
 \end{equation} 
For the parameter point under study, the predicted endpoint turns out to be 619 GeV if  we use the mass of the right-handed squarks, which  
are the ones that decay predominantly to ${\tilde \chi_1^0} q$.
 Figure \ref{fig:mt2} shows the $m_{T2}(0)$ distribution for  $~10$/fb of data, with the cuts of Table \ref{table:efficiencies} imposed. 
 A sharp drop-off leading up to $\sim 620$ GeV is evident, consistent with expectations. The spill-over to larger values is mostly due
 to the effects of extra jets not included in the calculation of the missing transverse energy  (in calculating  $M_{T2}(0)$ we use the missing transverse momentum as determined by the dijet system alone).

The  ($\alpha$, $\Delta \phi$) analysis we have described can be effective for higher-mass searches as well, 
with the cut on the sum of the two jets' transverse momenta  increased appropriately.  
Table \ref{table:summary}  gives results for other parameter points, with
the cuts on ${p_1}_T +{p_2}_T$ again chosen to give  robust values of $S/B$. The $M_0$ values are chosen to be near the lower bounds below which a stau LSP results. 
Provided that the squarks remain lighter than the gluino, increasing $M_0$ lowers the event rate somewhat 
but not dramatically.  For $(M_{1/2}, M_0)=(300,300)$ GeV, for example, the same cuts used for the $(M_{1/2}, M_0)=(300,100)$ GeV
point give 195 events after 1 fb$^{-1}$, with $S/B=4.7$.  

Taking $S/\sqrt{B}>5$ as the relevant criterion,  our results suggest that discovery through the dijet channel should
be possible for squark masses up t abut $\sim 1700$ GeV after 100 fb$^{-1}$ of integrated luminosity.  By the same measure,
discovery for lighter squark masses, $\sim 600$ GeV, should be possible after  $\sim$ a few$\times 10^2$ pb$^{-1}$ or less.  
It may be optimistic to focus on $S/\sqrt{B}$ as a discovery criterion, as doing so  assumes that the background is fully understood.  However, it is worth pointing out that (1) events with leptonic $Z$ decays will  provide some experimental handle on the dominant background,  $Z+$jets, and (2)
the shapes of the ${p_T}_1+{p_T}_2$ distributions for signal and background events passing the $\alpha$ and $\Delta \phi$ cuts are quite different (see Figure~\ref{fig:300ptbins}).  The excesses obtained in our analysis would lead to a prominent bump in the measured distribution, which would not be accommodated simply by  rescaling  the background.

%%%%%%%%%%%%%%%%%%%%%%%%%%%%%%%%%%%%%%%%%%%%%%%%%%%%%%%%%%%%%%%%%%%%%%%%%%%%%%%%%%%%%
\begingroup
\squeezetable
\begin{table}[t]
\begin{center}
\begin{tabular}{||c||c|c|c|c|c||}
\hline \hline
$(M_{1/2},\; M_0)$ & $(m_{\tilde g}, m_{\tilde q_R})$  &$\sum p_T$ cut & $\epsilon$ & $\sigma_{susy}$ (fb) & $S/B$ \\ \hline \hline
 (300, 100) & (716, 640)&  550 & $9.5\times 10^{-3}$ & 205. & 4.9 \\
  (450, 100) & (1040, 918)&  800 & $7.9  \times 10^{-3}$ & 21.3 & 4.7 \\
   (600, 150) & (1358, 1195)&  1050 & $8.1\times 10^{-3}$ & 4.07 & 5.0 \\
    (750, 200) & (1669, 1465)&  1250 & $9.6 \times 10^{-3}$ & 1.17& 4.8 \\
     (900, 200) & (1965, 1726)&  1450 & $1.0 \times 10^{-2}$ & 0.37 &  3.5 \\
\hline \hline
\end{tabular}
\end{center}
\label{default}
\caption{Efficiencies, event rates, and signal-to-background ratios for various SUSY parameters, using the cuts described in the text. All masses are in GeV.}
\label{table:summary}
\end{table}%
  \endgroup
  %%%%%%%%%%%%%%%%%%%%%%%%%%%%%%%%%%%%%%%%%%%%%%%%%%%%%%%%%%%%%%%%%%%%%%%%%%%%%%%%%%%%%

 \section{Conclusions}
We have studied several kinematic variables that can be used for dijet SUSY searches, and found that they give reasonable signal-to-background ratios.  Dijet events    can be used to constrain SUSY  mass parameters should the type of supersymmetry model we have considered be  correct.  Studies of  $Z+$jet events with leptonic $Z$ decays will give a better understanding of the background and a more reliable extraction of signal from background. For the future, it would be useful to see how well the lessons here can be applied to develop multijet searches that do not rely on full missing energy.

{\bf Acknowledgements} We wish to thank Patrick Meade and Liantao Wang for useful discussions in the early stages of this work. We also wish to thank  Maria Spiropulu, Patrick Janot, Oliver Buchmuller, Henning Flacher, and  the CMS SUSY analysis group for useful feedback and suggestions.  Finally,  we thank Ian Hinchliffe for bringing to our attention existing ATLAS dijet studies and ones in progress.    LR is supported by NSF grants PHY-0201124 and PHY-055611.  DTS is supported by NSF grant 0555421.
 
%%%%%%%%%%%%%%%%%%%%%%%%%%%%%%%%%%%%%%%%%%%%%%


\begin{thebibliography}{99}

\expandafter\ifx\csname natexlab\endcsname\relax\def\natexlab#1{#1}\fi
\expandafter\ifx\csname bibnamefont\endcsname\relax
  \def\bibnamefont#1{#1}\fi
\expandafter\ifx\csname bibfnamefont\endcsname\relax
  \def\bibfnamefont#1{#1}\fi
\expandafter\ifx\csname citenamefont\endcsname\relax
  \def\citenamefont#1{#1}\fi
\expandafter\ifx\csname url\endcsname\relax
  \def\url#1{\texttt{#1}}\fi
\expandafter\ifx\csname urlprefix\endcsname\relax\def\urlprefix{URL }\fi
\providecommand{\bibinfo}[2]{#2}
\providecommand{\eprint}[2][]{\url{#2}}



%\cite{:2007ww}
\bibitem{:2007ww}
  V.~M.~Abazov {\it et al.}  [D0 Collaboration],
  %``Search for squarks and gluinos in events with jets and missing transverse
  %energy using 2.1 fb-1 of ppbar collision data at sqrt(s)=1.96 TeV,''
  Phys.\ Lett.\  B {\bf 660}, 449 (2008)
  [arXiv:0712.3805 [hep-ex]].
  %%CITATION = PHLTA,B660,449;%%
  
\bibitem{atlas}  
ATLAS collaboration, ATLAS Physics TDR  CERN-LHCC 99-15;   
E. Richter-Was, D. Froidevaux, and J. Soderqvist,  ATLAS Internal Note ATL-PHYS-97-108 (1997);
M. Biglietti et. al., ATL-PHYS-2004-011;
 %\cite{Hinchliffe:1999zc}
%\bibitem{Hinchliffe:1999zc}
  I.~Hinchliffe and F.~E.~Paige,
  %``Measurements in SUGRA models with large tan(beta) at LHC,''
  Phys.\ Rev.\  D {\bf 61}, 095011 (2000)
  [arXiv:hep-ph/9907519];
  %%CITATION = PHRVA,D61,095011;%%
%\cite{Gianotti:2002xx}
%\bibitem{Gianotti:2002xx}
  F.~Gianotti {\it et al.},
  %``Physics potential and experimental challenges of the LHC luminosity
  %upgrade,''
  Eur.\ Phys.\ J.\  C {\bf 39}, 293 (2005)
  [arXiv:hep-ph/0204087].
  %%CITATION = EPHJA,C39,293;%%


\bibitem{cms} We thank O. Buchmuller, H. Flacher, J. Jones, T. Rommerskirchen and M. 
Stoye, as part of the CMS SUSY Analysis Group, for useful discussions and 
feedback.

\bibitem{ian} We thank Ian Hinchliffe for sharing this information.

%\cite{Hubisz:2008gg}
\bibitem{Hubisz:2008gg}
  J.~Hubisz, J.~Lykken, M.~Pierini and M.~Spiropulu,
  %``Missing energy look-alikes with 100 pb-1 at the LHC,''
  arXiv:0805.2398 [hep-ph].
  %%CITATION = ARXIV:0805.2398;%%


\bibitem[{\citenamefont{Djouadi
  et~al.}(2007{\natexlab{b}})\citenamefont{Djouadi, Muhlleitner, and
  Spira}}]{Djouadi:2006bz}
\bibinfo{author}{\bibfnamefont{A.}~\bibnamefont{Djouadi}},
  \bibinfo{author}{\bibfnamefont{M.~M.} \bibnamefont{Muhlleitner}},
  \bibnamefont{and} \bibinfo{author}{\bibfnamefont{M.}~\bibnamefont{Spira}},
  \bibinfo{journal}{Acta Phys. Polon.} \textbf{\bibinfo{volume}{B38}},
  \bibinfo{pages}{635} (\bibinfo{year}{2007}{\natexlab{b}}),
  \eprint{hep-ph/0609292}.

\bibitem[{\citenamefont{Sjostrand et~al.}(2006)\citenamefont{Sjostrand, Mrenna,
  and Skands}}]{Sjostrand:2006za}
\bibinfo{author}{\bibfnamefont{T.}~\bibnamefont{Sjostrand}},
  \bibinfo{author}{\bibfnamefont{S.}~\bibnamefont{Mrenna}}, \bibnamefont{and}
  \bibinfo{author}{\bibfnamefont{P.}~\bibnamefont{Skands}},
  \bibinfo{journal}{JHEP} \textbf{\bibinfo{volume}{05}}, \bibinfo{pages}{026}
  (\bibinfo{year}{2006}), \eprint{hep-ph/0603175}.


  
  \bibitem[{\citenamefont{Mangano et~al.}(2003)\citenamefont{Mangano, Moretti,
  Piccinini, Pittau, and Polosa}}]{Mangano:2002ea}
\bibinfo{author}{\bibfnamefont{M.~L.} \bibnamefont{Mangano}},
  \bibinfo{author}{\bibfnamefont{M.}~\bibnamefont{Moretti}},
  \bibinfo{author}{\bibfnamefont{F.}~\bibnamefont{Piccinini}},
  \bibinfo{author}{\bibfnamefont{R.}~\bibnamefont{Pittau}}, \bibnamefont{and}
  \bibinfo{author}{\bibfnamefont{A.~D.} \bibnamefont{Polosa}},
  \bibinfo{journal}{JHEP} \textbf{\bibinfo{volume}{07}}, \bibinfo{pages}{001}
  (\bibinfo{year}{2003}), \eprint{hep-ph/0206293}.
  
  
\bibitem[{PGS()}]{PGS}
\bibinfo{note}{The PGS simulation software is available at: \url
  {http://www.physics.ucdavis.edu/~conway/research/software/pgs/pgs4-general.h%
tm}.}

  

%\cite{Campbell:2003hd}
\bibitem{Campbell:2003hd}
  J.~Campbell, R.~K.~Ellis and D.~L.~Rainwater,
  %``Next-to-leading order QCD predictions for W + 2jet and Z + 2jet  production
  %at the CERN LHC,''
  Phys.\ Rev.\  D {\bf 68}, 094021 (2003)
  [arXiv:hep-ph/0308195].
  %%CITATION = PHRVA,D68,094021;%%
  
  \bibitem[{\citenamefont{Bonciani et~al.}(1998)\citenamefont{Bonciani, Catani,
  Mangano, and Nason}}]{Bonciani:1998vc}
\bibinfo{author}{\bibfnamefont{R.}~\bibnamefont{Bonciani}},
  \bibinfo{author}{\bibfnamefont{S.}~\bibnamefont{Catani}},
  \bibinfo{author}{\bibfnamefont{M.~L.} \bibnamefont{Mangano}},
  \bibnamefont{and} \bibinfo{author}{\bibfnamefont{P.}~\bibnamefont{Nason}},
  \bibinfo{journal}{Nucl. Phys.} \textbf{\bibinfo{volume}{B529}},
  \bibinfo{pages}{424} (\bibinfo{year}{1998}), \eprint{hep-ph/9801375}.
  
  \bibitem[{pro()}]{prospino}
\bibinfo{note}{Prospino 2.0 is available at:
  \url{http://www.ph.ed.ac.uk/~tplehn/prospino/}.}

\bibitem[{\citenamefont{Beenakker et~al.}(1997)\citenamefont{Beenakker, Hopker,
  Spira, and Zerwas}}]{Beenakker:1996ch}
\bibinfo{author}{\bibfnamefont{W.}~\bibnamefont{Beenakker}},
  \bibinfo{author}{\bibfnamefont{R.}~\bibnamefont{Hopker}},
  \bibinfo{author}{\bibfnamefont{M.}~\bibnamefont{Spira}}, \bibnamefont{and}
  \bibinfo{author}{\bibfnamefont{P.~M.} \bibnamefont{Zerwas}},
  \bibinfo{journal}{Nucl. Phys.} \textbf{\bibinfo{volume}{B492}},
  \bibinfo{pages}{51} (\bibinfo{year}{1997}), \eprint{hep-ph/9610490}.


%\cite{Lester:1999tx}
\bibitem{Lester:1999tx}
  C.~G.~Lester and D.~J.~Summers,
%   ``Measuring masses of semi-invisibly decaying particles pair produced at
  %hadron colliders,''
  Phys.\ Lett.\  B {\bf 463}, 99 (1999)
  [arXiv:hep-ph/9906349];
  %%CITATION = PHLTA,B463,99;%%
%\cite{Barr:2003rg}
%\bibitem{Barr:2003rg}
  A.~Barr, C.~Lester and P.~Stephens,
  %``m(T2): The truth behind the glamour,''
  J.\ Phys.\ G {\bf 29}, 2343 (2003)
  [arXiv:hep-ph/0304226].
  %%CITATION = JPHGB,G29,2343;%%

\end{thebibliography}
\end{document}